\documentstyle[preprint,aps,epsfig,floats]{revtex}

\newcommand{\ac}{{\centerline{\bf Acknowledgments}}}
\tightenlines
\draft
\begin{document}
\date{\today}
\preprint{\vbox{\hbox{WUB 99-6}\hbox{RUB-TPII-03/99}}}
\title{Large-order trend of the anomalous-dimensions spectrum of
       trilinear twist-3 quark operators\\}
\author{M. Bergmann,${}^{1}$\thanks{E-mail:
        michael.bergmann@sercon.de}
        W. Schroers,${}^{2}$\thanks{Email:
        wolfram@theorie.physik.uni-wuppertal.de}
        and
        N.\ G.\ Stefanis,${}^{3}$\thanks{Email:
        stefanis@tp2.ruhr-uni-bochum.de}
        }
\address{${}^{1}$ SerCon, c/o IBM Deutschland GmbH,         \\
                  Karl-Arnold-Platz 1A,                     \\
                  D-40774 D\"usseldorf, Germany             \\
                  [0.3cm]
         ${}^{2}$ Fachbereich Physik,                       \\
                  Universit\"at Wuppertal,                  \\
                  D-42097 Wuppertal, Germany                \\
                  [0.3cm]
         ${}^{3}$ Institut f\"ur Theoretische Physik II,    \\
                  Ruhr-Universit\"at Bochum,                \\
                  D-44780 Bochum, Germany                   \\
         }
\maketitle
\begin{abstract}
The anomalous dimensions of trilinear-quark operators are calculated
at leading twist $t=3$ by diagonalizing the one-gluon exchange kernel
of the renormalization-group type evolution equation for the nucleon
distribution amplitude. This is done within a symmetrized basis of
Appell polynomials of maximum degree $M$ for $M\gg 1$ (up to
order 400) by combining analytical and numerical algorithms. The
calculated anomalous dimensions form a degenerate system, whose
upper envelope shows asymptotically logarithmic behavior.
\end{abstract}
\newpage       
\input amssym.def
\input amssym.tex
The anomalous dimensions of field operators are crucial ingredients in
determining the scale dependence of hadronic wave functions in QCD.
Though the calculation of the bound-state wave function at the initial
(low-momentum) scale cannot be determined within perturbative QCD and
requires nonperturbative methods (for a review, see, \cite{CZ84}), its
evolution is governed by a renormalization-group type evolution
equation \cite{LB80} which in leading order reduces to the
one-gluon-exchange kernel.

Bilinear quark operators of lowest twist ($t=2$) are representations of
the collinear conformal group and therefore their anomalous dimensions
(to first order in $\alpha _{\rm s}$) are completely determined by
collinear conformal invariance (see, e.g.,
\cite{ER80,Bro80,Ohr81,Mue94}).
Therefore, the eigenvalue problem of twist-2 quark operators with an
arbitrary number of total derivatives is exactly solvable.
In contrast, the anomalous-dimensions matrix of trilinear quark
operators of leading twist-3 cannot be diagonalized completely by
collinear conformal invariance
\cite{Pes79,Kre79,Tes82,Ohr82,Nye92,THESIS97}.
As a result, neither the eigenfunctions nor the eigenvalues
(alias the anomalous dimensions) of the nucleon evolution equation are
known explicitly.
Hence one has to diagonalize the anomalous dimensions matrix order by
order within an appropriate operator basis.

While operators involving total derivatives do not contribute to
forward scattering amplitudes, like deep-inelastic scattering,
exclusive processes receive contributions from non-forward matrix
elements, and consequently the mixing with trilinear quark operators
of twist-3 containing total derivatives cannot be neglected.
Given that the baryon (hadron) distribution amplitudes are determined
at low normalization scales, of the order of 1~GeV or less,
\cite{CZ84,Ste89,BS93,PPRWG98}, the knowledge of anomalous dimensions
is indispensable in confronting calculated observables, like the proton
form factor, with experimental data at larger values of the momentum
transfer.

In this letter we present a systematic treatment of the nucleon
evolution equation, focusing on the large-order, i.e., quasi-asymptotic,
behavior of the spectrum of the anomalous dimensions of twist-3
multiplicatively renormalizable baryonic $I_{1/2}$-operators with the
number $M$ of total derivatives acting on these fields tending to
infinity.
More precisely, we consider operators of the following form
\begin{equation}
  {\cal V}_{\gamma}^{\left( n_{1}n_{2}n_{3} \right)}(0)
\equiv
  \left( iz \cdot D \right)^{n_{1}} u(0)
  \left[ C \, \gamma _{\mu} \, z_{\mu} \right]
  \left( iz \cdot D \right)^{n_{2}} u(0) \,
  \left( iz \cdot D \right)^{n_{3}}
  \gamma _{5} d_{\gamma}(0)
\label{eq:Vder}
\end{equation}
with $n_{1} + n_{2} + n_{3} = M$, and where $z$ is an auxiliary
lightlike vector (i.e., $z^{2} = 0$) to project out the leading twist
contribution \cite{CZ84}.
Note that since there are $M+1$ independent operators at each order, one
has to diagonalize a $(M+1)\times(M+1)$ matrix in order to determine the
corresponding eigenvectors and eigenvalues.
This is done here within a basis of trilinear quark operators in terms
of appropriately symmetrized Appell polynomials (see below) by
analytically diagonalizing the nucleon evolution equation up to order
7, and proceeding beyond that order up to a maximum order of 400
numerically.

We have already elaborated on these issues in a series of papers
\cite{ELBA93,KYF93,Ber94,COMO94,ZAKO94,THESIS97}, and the present work
complements and generalizes our previous results, establishing in
particular that the upper envelope of the spectrum for very large order
$M$ follows a logarithmic law given by the fit
\begin{equation}
    \gamma _{n}(M)
=
  - 0.638 + 0.8882 \ln (M + 1.91)
\label{eq:logfit}
\end{equation}
for any member of the spectrum labeled by the index $n$.

Before proceeding with the presentation of our analysis, let us first
mention a very recent alternative approach by Braun and collaborators
\cite{BDKM99}, based on the observation that the eigenvalues of a
helicity $\lambda = 3/2$ baryon, (say, the $\Delta ^{++}$) can be
determined exactly because the corresponding Hamiltonian
${\cal H}_{3/2}$ is integrable and can be formally treated as a
one-dimensional XXX Heisenberg spin magnet of noncompact spin $s=-1$.
Though the Hamiltonian ${\cal H}_{1/2}$ (relevant for the nucleon) is
not integrable and consequently the corresponding eigenvalue problem
is not exactly sovable, these authors argue that the helicity-flip
terms can be treated as a small perturbation, so that the spectrum of
anomalous dimensions can be determined by calculating the mixing matrix
elements of these terms in the basis of the exact eigenfunctions of
${\cal H}_{3/2}$.
This is an attractive mathematical framework, and we shall compare our
results with theirs at the end of the paper.

Let us now enter our approach in more detail.

The evolution equation is given by \cite{LB80}
\begin{equation}
  x_{1}x_{2}x_{3} \,
  \left[
          \frac{\partial}{\partial \xi}{\tilde\Phi}
          \left( x_{i}, Q^{2} \right)
        + \frac{3}{2} \frac{C_{\rm F}}{\beta _{0}} \,
          \tilde{\Phi}\left (x_{i}, Q^{2} \right)
  \right]
=
  \frac{C_{\rm F}}{\beta _{0}} \,
  \int_{0}^{1}[dy] \,
  V\left( x_{i}, y_{i} \right)\,
  \tilde{\Phi}\left( y_{i}, Q^{2} \right) \; ,
\label{eq:evolequ}
\end{equation}
where
\begin{equation}
  \xi
\equiv
  \frac{\beta _{0}}{4\pi}
  \int_{\mu ^{2}}^{Q^{2}} \frac{dk_{\perp}^{2}}{k_{\perp}^{2}} \,
                   \alpha _{\rm s}(k_{\perp}^{2})
=
  \ln \frac{\alpha _{\rm s}(\mu ^{2})}{\alpha _{\rm s}(Q^{2})}
=
  \ln \frac{\ln Q^{2}/\Lambda _{{\rm QCD}}^{2}}
           {\ln \mu ^{2}/\Lambda _{{\rm QCD}}^{2}}
\label{eq:xi}
\end{equation}
is the evolution ``time'' parameter,
$
 \beta _{0}
=
 11-\frac{2}{3}n_{\rm f}
=
 9
$ for three flavors,
$\Phi = x_{1}x_{2}x_{3} \, \tilde{\Phi}$
denotes the full nucleon distribution amplitude, $\{\tilde{\Phi}\}$
its eigenfunctions, and
$
 C_{\rm F}
=
 \left( N_{\rm c}^{2} - 1 \right)/2N_{\rm c}
$,
with the integration measure defined by
$
 [dy]
=
 dy_{1}dy_{2}dy_{3} \delta \left( 1 - y_{1} - y_{2} - y_{3} \right)
$.
The kernel
$
 V\left( x_{i}, y_{i} \right)
=
 V\left( y_{i}, x_{i} \right)
$
is the sum over one-gluon interactions between quark pairs $\{ i,j \}$
at order $\alpha _{\rm s}$ and is not a function but a distribution from
which infrared divergences at $x_{i} = y_{i}$ have been removed.

We propose to solve this evolution equation by employing factorization
of the dependence on longitudinal momentum fractions from that on the
external (large) momentum scale $Q^{2}$, the latter being controlled by
\begin{equation}
    \frac{\partial}{\partial \xi}\,
    \tilde{\Phi}_{n}\left( x_{i}, Q^{2} \right)
=
  - \gamma _{n}\, \tilde{\Phi}_{n}\left( x_{i}, Q^{2} \right)
\label{eq:eivaleq}
\end{equation}
with solutions
\begin{equation}
  \tilde{\Phi}_{n}\left( x_{i}, Q^{2} \right)
=
  \tilde{\Phi}_{n}\left( x_{i} \right)
\left[
      \frac{\alpha _{\rm s} \left( Q^{2} \right)}
           {\alpha _{\rm s} \left( \mu ^{2} \right)}
\right]^{\gamma _{n}}
\simeq
  \tilde{\Phi}_{n}\left( x_{i} \right)
\left(
      \ln \frac{Q^{2}}{\Lambda _{\rm QCD}^{2}}
\right)^{- \gamma _{n}} \; .
\label{eq:nueifun}
\end{equation}
This allows us to write the full nucleon distribution amplitude in the
form
\begin{equation}
  \Phi \left( x_{i}, Q^{2} \right)
\sim
  x_{1}x_{2}x_{3} \,
  \sum_{n = 0}^{\infty} B_{n}\left(\mu ^{2}\right) \,
  \tilde{\Phi}_{n}\left( x_{i} \right)\,
  \left(
        \ln \frac{Q^{2}}{\Lambda _{\rm QCD}^{2}}
  \right)^{- \gamma _{n}} \; ,
\label{eq:solnucevol}
\end{equation}
where $\tilde{\Phi}_{n}(x_{i})$ are appropriate but not tabulated
polynomials, and the expansion coefficients $B_{n}$ encode the
nonperturbative input of the bound-states dynamics at the
factorization scale $\mu$.

From the factorized form of
$
 \tilde{\Phi}_{n}\left( x_{i}, Q^{2} \right)
$
in Eq.~(\ref{eq:nueifun}), it follows that the evolution equation
for the $x$-dependence reduces to the characteristic equation
\begin{equation}
  x_{1}x_{2}x_{3} \,
  \left[
          \frac{3}{2}\frac{C_{\rm F}}{\beta _{0}} \,
        - \gamma _{n}
  \right] \,
          \tilde{\Phi}\left( x_{i} \right)
=
  \frac{C_{\rm B}}{\beta _{0}} \,
  \int_{0}^{1} [dy] \,
  V\left( x_{i}, y_{i} \right)
         \tilde{\Phi}\left( y_{i} \right)
\label{eq:evolxpart}
\end{equation}
with
$
 C_{\rm B}
=
 \left( N_{\rm c} + 1 \right)/2N_{\rm c}
$.

To proceed, it is convenient to consider the kernel
$V\left( x_{i}, y_{i} \right)$ as being an operator expanded on the
polynomial basis~\cite{LB80}
$
 |x_{1}^{k}\, x_{3}^{l} \rangle
\equiv
 |k\,l\rangle
$
(recall that because of momentum conservation, only two out of three
$x_{i}$ variables are linearly independent), i.e., to write
\begin{equation}
  \hat{V}
\equiv
  \int_{0}^{1}[dy]\, V\left( x_{i}, y_{i} \right)
\label{eq:voperx1x3}
\end{equation}
and convert Eq.~(\ref{eq:evolxpart}) into the algebraic equation
\begin{equation}
  \left[
            \frac{3}{2}\frac{C_{\rm F}}{\beta _{0}} \,
        - 2 \frac{C_{\rm B}}{\beta _{0}} \,
            \frac{\hat{V}}{2w\left( x_{i} \right)}
  \right]
  \tilde{\Phi}_{n}\left( x_{i} \right)
=
  \gamma _{n} \, \tilde{\Phi}_{n}\left( x_{i} \right) \; ,
\label{eq:algnucevoleq}
\end{equation}
where
$
 w\left( x_{i} \right)
=
 x_{1}x_{2}x_{3} = x_{1}\left( 1 - x_{1} - x_{3} \right)x_{3}$
is the weight function of the orthogonal basis.
In this way, the action of the operator $\hat{V}$ can be completely
determined by a matrix, namely:
\begin{equation}
  \frac{\hat{V}|k \, l \rangle}{2 w\left( x_{i} \right)}
=
  \frac{1}{2} \sum_{i,j}^{i + j\leq M} |i\, j \rangle U_{ij,kl} \; .
\label{eq:kernux1x3}
\end{equation}
The corresponding eigenvalues are then determined by the roots
$\eta _{n}$ of the characteristic polynomial that diagonalizes the
matrix $U$:
\begin{equation}
    \hat{V} \, \tilde{\Phi}_{n}\left( x_{i} \right)
=
  - \eta _{n} \,
    w\left( x_{i} \right) \,
    \tilde{\Phi}_{n}\left( x_{i} \right) \; ,
\end{equation}
\label{eq:rootsVhat}
so that the anomalous dimensions are given by
\begin{equation}
 \gamma _{n}(M)
=
 \frac{1}{\beta_{0}}
 \left(
       \frac{3}{2}C_{\rm F} + 2\eta _{n}(M)C_{\rm B}
 \right) \; ,
\label{eq:gamma_n}
\end{equation}
with the orthogonalization prescription
\begin{equation}
  \int_{0}^{1}[dx] \, w\left( x_{i} \right) \,
  \tilde{\Phi}_{m}\left( x_{i} \right)
  \tilde{\Phi}_{n}\left( x_{i} \right)
=
  \frac{1}{N_{m}}\, \delta _{mn} \; ,
\label{eq:orthonuc}
\end{equation}
where $N_{m}$ are appropriate normalization constants (see Table
\ref{tab:eigen}).

Within the basis $|k\, l\rangle$, the matrix $U$ can be diagonalized
to provide eigenfunctions, which are polynomials of degree
$M = k + l = 0,1,2,3 \ldots$, with $M + 1$ eigenfunctions for each $M$.
This will be done within a basis of symmetrized Appell polynomials
defined by \cite{Ber94,ZAKO94,THESIS97}
\begin{eqnarray}
  \tilde{{\cal F}}_{mn}\left( x_{1}, x_{3} \right)
& = &
  \frac{1}{2}\left[ {\cal F}_{mn}\left( x_{1}, x_{3} \right)
\pm
  {\cal F}_{nm}\left( x_{1}, x_{3} \right) \right]
\nonumber \\
& = &
  \sum_{k,l = 0}^{k + l\leq m + n} Z_{kl}^{mn} |k\, l\rangle \; ,
\label{eq:symApp}
\end{eqnarray}
where $+$ refers to the case $m\geq n$ and $-$ to the case $m<n$, and
the Appell polynomials are defined in terms of special hypergeometric
functions, according to
\begin{equation}
  {\cal F}_{mn}(x_{1}, x_{3})
\equiv
  {\cal F}_{mn}^{(M)}(5, 2, 2; x_{1}, x_{3}) \; .
\label{eq:appellpol}
\end{equation}

These polyomials constitute an orthogonal set on the triangle
$T = T(x_{1},x_{3})$ with $x_{1} > 0$, $x_{3} > 0$, $x_{1} + x_{3} < 1$
and provide a very suitable basis to solve the eigenvalue equation
for the nucleon because $\hat{V}$ (i) is blockdiagonal for different
polynomial orders, and (ii) commutes with the permutation operator
$P_{13}=[321]$, i.e.,
$
 [P_{13},\hat{V}]
=
 0
$.
Hence $\hat{V}$ becomes, in addition, blockdiagonal within each sector
of permutation-symmetry eigenfunctions at fixed order $M$.
As a result, the kernel $\hat{V}$ can be analytically diagonalized up
to order seven, providing a total of
$n_{\rm max}(M)
 =
 \frac{1}{2}(M+1)(M+2)
=
36
$
eigenvectors and eigenvalues (alias eigenfunctions and anomalous
dimensions).
Beyond that order, the evolution equation has to be solved numerically
because the roots of the characteristic polynomial of matrices with rank
four cannot be determined analytically.

From the practical point of view, such a large set of eigenfunctions
surely exceeds the number of theoretical constraints on the
nonperturbative input coefficients $B_{n}$ that can be derived from
QCD sum rules \cite{CZ84,KS87,COZ89}, lattice simulations \cite{RSS87},
or by fitting experimental data \cite{BK96}.
However, yet much larger orders are necessary in order to extract the
asymptotic behavior of the anomalous-dimensions spectrum.

For this purpose, let us express all nucleon eigenfunctions as linear
combinations of symmetrized Appell polynomials of the same order $M$
\begin{equation}
  \tilde{\Phi}_{k}\left( x_{i} \right)
=
  \sum_{m,n=0}^{m+n=M} \,
  c_{mn}^{k}{\cal F}_{mn}\left( 5, 2, 2; x_{1}, x_{3} \right)
\label{eq:Appellphi}
\end{equation}
and rearrange $\tilde{{\cal F}}_{mn}$ (cf. Eq. (\ref{eq:symApp})), which
belongs to a definite symmetry class $S_{n}=\pm 1$ within order $M$, in
the form of an (arbitrary) vector to read
\begin{equation}
  \tilde{{\cal F}}_{mn}\left( x_{1}, x_{3} \right)
\longmapsto
  \tilde{{\cal F}}_{q}\left( x_{1}, x_{3} \right) \; .
\label{eq:symAppvec}
\end{equation}
Then Hilbert-Schmidt orthogonalization yields a basis
\begin{equation}
  |\tilde{{\cal F}}_{q}^{\prime} \rangle
=
  \sum_{k,l}^{}Z_{kl}^{q}|k \, l \rangle
\label{eq:orthobasis}
\end{equation}
with
\begin{equation}
  \int_{0}^{1} [dx] \, w\left( x_{i} \right) \,
  \tilde{{\cal F}}_{q}^{\prime}
  \tilde{{\cal F}}_{q^{\prime}}^{\prime}
\propto
  \delta _{qq^{\prime}} \; ,
\label{eq:orthorel}
\end{equation}
so that
\begin{equation}
  \frac{\hat{V}|\tilde{{\cal F}}_{q}^{\prime} \rangle}
       {2w\left( x_{i} \right)}
=
  \frac{1}{2}
  \sum_{i,j,k,l}^{}Z_{kl}^{q}\, U_{ij,kl} |i \, j \rangle \; .
\label{eq:hatVinF_q}
\end{equation}
Note that the construction of polynomials depending on two variables
via the Hilbert-Schmidt method has no unique solution, but depends on
the order in which the orthogonalization is performed.
Since beyond order $M=3$, neither the eigenvalues nor the
normalization factors are rational numbers, one has to find which
representation is more convenient for calculations.

The last step in determining the eigenfunctions and eigenvalues of
$\hat{V}$ is to define the matrix
\begin{equation}
  {\cal M}_{q^{\prime}q}
=
  \int_{0}^{1}[dx]\,w\left( x_{1}, (1-x_{1}-x_{3}), x_{3} \right)
  \tilde{{\cal F}}_{q^{\prime}}^{\prime}\left( x_{1}, x_{3} \right)
  \hat{V}\left( x_{1}, x_{3} \right)
  \tilde{{\cal F}}_{q}^{\prime}\left( x_{1}, x_{3} \right)
\label{eq:eigenmatrix}
\end{equation}
and calculate the roots of the characteristic polynomial
\begin{equation}
  {\cal P}(\eta )
=
  \det \left[{\cal M}_{q^{\prime}q} - \eta I_{q^{\prime}q}\right] \; .
\label{eq:P}
\end{equation}
Consequently, in terms of the eigenvectors
$
 \bbox{m}_{q}
=
\left( m_{q}^{1}, \ldots m_{q}^{q^{\prime}} \right)
$
of ${\cal M}_{q^{\prime}q}$,
the eigenfunctions of the evolution equation are given by
\begin{eqnarray}
  \tilde{\Phi}_{q}\left( x_{1}, x_{3} \right)
& \propto &
  \sum_{q^{\prime}}^{}m_{q}^{q^{\prime}}
  \tilde{{\cal F}}_{q^{\prime}}^{\prime}\left( x_{1}, x_{3}\right)
\nonumber \\
& = &
  \sum_{k,l}^{} a_{kl}^{q}\, |k \, l \rangle \; .
\label{eq:eigenfuns}
\end{eqnarray}
For every order $M$, there are $M+1$ eigenfunctions of the same order
with an excess of symmetric terms by one for even orders.
A compedium of the results up to order $M=4$, yielding a total of
15 eigenfunctions and associated anomalous dimensions, is given in
Table~\ref{tab:eigen}.
The precision of orthogonality is at least $10^{-8}$.

\begin{table}
\squeezetable
\caption{Orthogonal eigenfunctions
         $
            \tilde{\Phi}_n\left( x_{i} \right)
          = \sum_{kl}^{} \,
            a_{kl}^{n} \,
            x_{1}^{k}x_{3}^{l}
         $
         of the nucleon evolution equation up to order $M=4$ in terms
         of the coefficient matrix
         $a_{kl}^{n}$ ($a_{kl}^{n} = S_{n}\, a_{lk}^{n}$ with $n$
         fixed) and the corresponding anomalous dimensions
         $\gamma _{n}(M)$ defined in the text.
         The numerical results for $n\geq 12$ have been obtained with
         a much higher numerical accuracy than shown in this table.
\label{tab:eigen}}
\begin{tabular}{rc|rcccc}
$ n  $&$ M $&$ S_n $&$\gamma_n$&$\eta_n$&$ N_n$&$a_{00}^{n}$\\
\hline
 $ 0$&$ 0$&$1$&${2\over {27}}$&$-1$&$120$&$1$\\
 $ 1$&$ 1$&$-1$&${{26}\over {81}}$&${2\over 3}$&$1260$&$0$\\
 $ 2$&$ 1$&$1$&${{10}\over {27}}$&$1$&$420$&$-2$\\
 $ 3$&$ 2$&$1$&${{38}\over {81}}$&${5\over 3}$&$756$&$2$\\
 $ 4$&$ 2$&$-1$&${{46}\over {81}}$&${7\over 3}$&$34020$&$0$\\
 $ 5$&$ 2$&$1$&${{16}\over {27}}$&${5\over 2}$&$1944$&$2$\\
 $ 6$&$ 3$&$1$&${{115 - {\sqrt{97}}}\over {162}}$&${{-\left( -79 + {\sqrt{97}} \right) }\over {24}}$&${{4620\,\left( 485 + 11\,{\sqrt{97}} \right) }\over {97}}$&$1$\\
 $ 7$&$ 3$&$1$&${{115 + {\sqrt{97}}}\over {162}}$&${{79 + {\sqrt{97}}}\over {24}}$&${{4620\,\left( 485 - 11\,{\sqrt{97}} \right) }\over {97}}$&$1$\\
 $ 8$&$ 3$&$-1$&${{559 - {\sqrt{4801}}}\over {810}}$&${{-\left( -379 + {\sqrt{4801}} \right) }\over {120}}$&${{27720\,\left( 33607 - 247\,{\sqrt{4801}} \right)}\over {4801}}$&$0$\\
 $ 9$&$ 3$&$-1$&${{559 + {\sqrt{4801}}}\over {810}}$&${{379 + {\sqrt{4801}}}\over {120}}$&${{27720\,\left( 33607 + 247\,{\sqrt{4801}} \right) }\over {4801}}$&$0$\\
 $ 10$&$4$&$-1$&${{346 - {\sqrt{1081}}}\over {405}}$&${{-\left( -256 + {\sqrt{1081}} \right) }\over {60}}$&${{196560\,\left( 7567 - 13\,{\sqrt{1081}} \right) }\over {1081}}$&$0$\\
 $ 11$&$4$&$-1$&${{346 + {\sqrt{1081}}}\over {405}}$&${{256 + {\sqrt{1081}}}\over {60}}$&${{196560\,\left( 7567 + 13\,{\sqrt{1081}} \right)}\over{1081}}$&$0$\\
 $ 12$&$4$&$1$&$0.70204$&$3.23876$&$1$&$153.37061$ \\
 $ 13$&$4$&$1$&$0.80651$&$3.94397$&$1$&$332.500864$ \\
 $ 14$&$4$&$1$&$0.93589$&$4.81727$&$1$&$-137.11538$
\end{tabular}
\vspace{-7 pt}
\begin{tabular}{r|cccccccc}
$ n  $&$a_{10}^n$&$a_{20}^n$&$a_{11}^n$&$a_{30}^n$&
       $a_{21}^n$&$a_{40}^n$&$a_{31}^n$&$a_{22}^n$ \\
\hline
 $  0$&$0$&$0$&$0$&$0$&$0$&$0$&$0$&$0$\\
 $  1$&$1$&$0$&$0$&$0$&$0$&$0$&$0$&$0$\\
 $  2$&$3$&$0$&$0$&$0$&$0$&$0$&$0$&$0$\\
 $  3$&$-7$&$8$&$4$&$0$&$0$&$0$&$0$&$0$\\
 $  4$&$1$&$-{4\over 3}$&$0$&$0$&$0$&$0$&$0$&$0$\\
 $  5$&$-7$&${{14}\over 3}$&$14$&$0$&$0$&$0$&$0$&$0$\\
 $  6$&$-6$&${{41 + {\sqrt{97}}}\over 4}$&${{3\,\left( 31 - {\sqrt{97}} \right) }\over 4}$&${{-5\,\left( 17 + {\sqrt{97}} \right) }\over {16}}$&${{-5\,\left( 31 - {\sqrt{97}} \right) }\over 8}$&$0$&$0$&$0$\\
 $  7$&$-6$&${{41 - {\sqrt{97}}}\over 4}$&${{3\,\left( 31 + {\sqrt{97}} \right) }\over 4}$&${{-5\,\left( 17 - {\sqrt{97}} \right) }\over {16}}$&${{-5\,\left( 31 + {\sqrt{97}} \right) }\over 8}$&$0$&$0$&$0$\\
 $  8$&$1$&$-3$&$0$&${{601 + {\sqrt{4801}}}\over {264}}$&${{59 - {\sqrt{4801}}}\over {44}}$&$0$&$0$&$0$\\
 $  9$&$1$&$-3$&$0$&${{601 - {\sqrt{4801}}}\over {264}}$&${{59 + {\sqrt{4801}}}\over {44}}$&$0$&$0$&$0$\\
 $ 10$&$1$&$-5$&$0$&${{379 + {\sqrt{1081}}}\over {48}}$&${{61 - {\sqrt{1081}}}\over 8}$&${{-\left( 159 + {\sqrt{1081}} \right) }\over {40}}$&${{-\left( 61 - {\sqrt{1081}} \right) }\over 8}$&$0$\\
 $ 11$&$1$&$-5$&$0$&${{379 - {\sqrt{1081}}}\over {48}}$&${{61 + {\sqrt{1081}}}\over 8}$&${{-\left( 159 - {\sqrt{1081}} \right) }\over {40}}$&${{-\left( 61 + {\sqrt{1081}} \right) }\over 8}$&$0$\\

$12$&$-1380.33552$&$5232.86956$&$5006.42414$&$-8063.85349$&$-9178.44426$&$4345.63139$&$4926.80699$&$8503.27454$\\

$13$&$-2992.50778$&$9240.51876$&$17166.06044$&$-11695.76593$&$-31471.11081$&$5068.49438$&$19489.65169$&$23962.91822$\\

$14$&$1234.03849$&$-1843.05428$&$-12981.41464$&$-587.61051$&$23799.26017$&$1382.85660$&$-10302.90296$&$-26992.71442$\\
\end{tabular}
\end{table}

It turns out that the eigenfunctions $\{\tilde{\Phi}_{n}\}$
of the nucleon evolution equation satisfy a commutative algebra
subject to the triangular condition
$
 |{\cal O}(k) - {\cal O}(l)| \leq {\cal O}(m) \leq {\cal O}(k) +
 {\cal O}(l):
$
\begin{equation}
  \tilde{\Phi}_{k}\left( x_{i} \right)
  \tilde{\Phi}_{l}\left( x_{i} \right)
=
  \sum_{m=0}^{\infty}\,F_{kl}^{m}\tilde{\Phi}_{m}\left( x_{i} \right)
\label{eq:algebra}
\end{equation}
with structure coefficients $F_{kl}^{m}$ given by
\begin{equation}
  F_{kl}^{m}
=
  N_{m} \int_{0}^{1}[dx]\, x_{1}x_{3}(1-x_{1}-x_{3})
  \tilde{\Phi}_{m}\left( x_{i} \right)
  \tilde{\Phi}_{k}\left( x_{i} \right)
  \tilde{\Phi}_{l}\left( x_{i} \right) \; ,
\label{eq:structurecoefficients}
\end{equation}
${\cal O}(k)$ being defined by
\begin{equation}
  {\cal O}(k) = \left\{
\begin{array}{ll}
       0  & \quad\quad k = 0 \cr
       1  & \quad\quad 1 \leq k \leq 2 \cr
       2  & \quad\quad 3 \leq k \leq 5 \cr
       3  & \quad\quad 6 \leq k \leq 9 \cr
       4  & \quad\quad k = 10,11       \cr
  \vdots  & \quad\quad \ldots
\end{array} \right.
\label{eq:counter}
\end{equation}
Note that the structure coefficients are symmetric, i.e.,
$
 F_{kl}^{m} = F_{lk}^{m}.
$
Furthermore, $F_{kk}^{0}=\frac{N_{0}}{N_{k}}$.
The utility of this algebra derives from the fact that once
the structure coefficients have been computed, they can be used
to express any function $f(x_{1},x_{3})$ in terms of the nucleon
eigenfunctions.
The values of $F_{lk}^{m}$ up to ${\cal O}(k)=11$ are tabulated
in \cite{Ber94}.
Unfortunately, it is not that easy to identify the proper group (and
physical symmetry) underlying this algebra.
Work in this direction is still in progress.

\begin{figure}
\tighten
\[
\psfig{figure=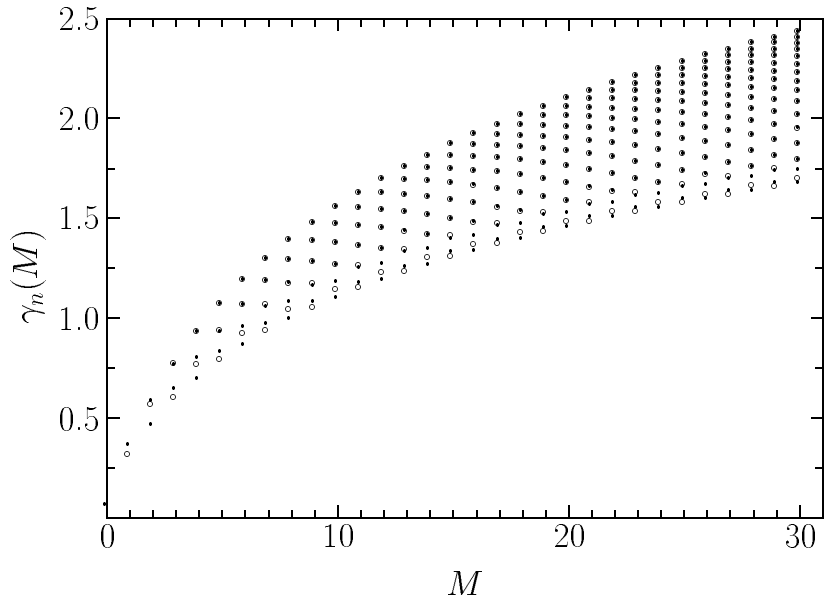,%
       bbllx=96pt,bblly=390pt,bburx=485pt,bbury=685pt,%
       width=10.5cm,silent=}
\]
\caption[fig:spectrum30]
        {\tenrm Illustration of the spectrum of the anomalous dimensions
         up to order $M=30$, comprising symmetric (full circles) and
         antisymmetric (open circles) contributions.
\label{fig:spec30}}
\end{figure}
%

Next, we present the results obtained for the anomalous-dimensions
spectrum.
An impression of the detailed structure of the spectrum is provided
by fig. \ref{fig:spec30}.
Both symmetry classes under the permutation $P_{13}$ are shown:
open circles stand for values belonging to $S_{n}=-1$ (antisymmetric
sector) and black dots for those belonging to $S_{n}=1$ (symmetric
sector).
As the order $M$ increases, the degeneracy of the spectrum also
increases because more and more operators with the same quantum
numbers contribute and the density of eigenvalues becomes very high.
Because all $\gamma _{n}$ are positive fractional numbers increasing
with the counting index $n$, higher terms in the eigenfunctions
decomposition of the nucleon distribution amplitude are gradually
suppressed (cf. Eq. \ref{eq:solnucevol}).

\begin{figure}
\tighten
\[
\psfig{figure=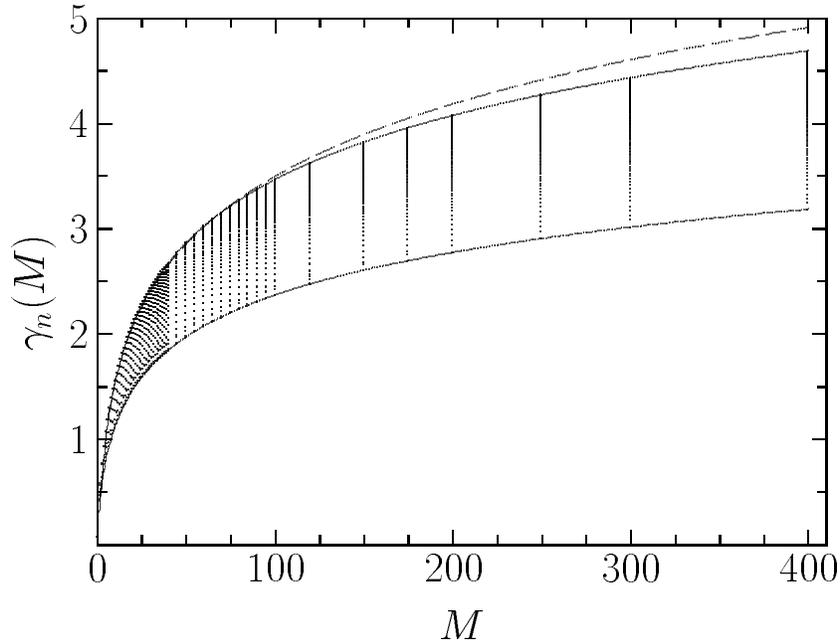,%
       bbllx=88pt,bblly=426pt,bburx=419pt,bbury=684pt,%
       width=11.0cm,silent=}
\]
\caption[fig:spectrum400]
        {\tenrm Spectrum of the anomalous dimensions of trilinear
         twist-3 quark operators up to order $M=400$. The solid lines
         (upper and lower envelopes of the spectrum) represent
         logarithmic fits up to the maximum considered order 400, taking
         into consideration all orders above 10.
         The dashed line gives for comparison a previous logarithmic fit
         \cite{THESIS97} (see text) which takes into account all orders
         up to 150.
\label{fig:spec400}}
\end{figure}

At very large order a different picture for the large-order behavior of
the spectrum of anomalous dimensions develops, namely one of logarithmic
rise.
Indeed, while the low-order spectrum with $M\leq 3$
\cite{LB80,Pes79,Kre79,Tes82} can be reproduced by the empirical
power-law
$
 \gamma _{n}(M)
=
 0.37 M^{0.565}
$,
the inclusion of large orders seems to be better described by a
logarithm.
Extending the calculation up to a maximum order of $M=400$, we finally
obtain the spectrum displayed in fig. \ref{fig:spec400}.

%

\begin{figure}
\tighten
\[
\psfig{figure=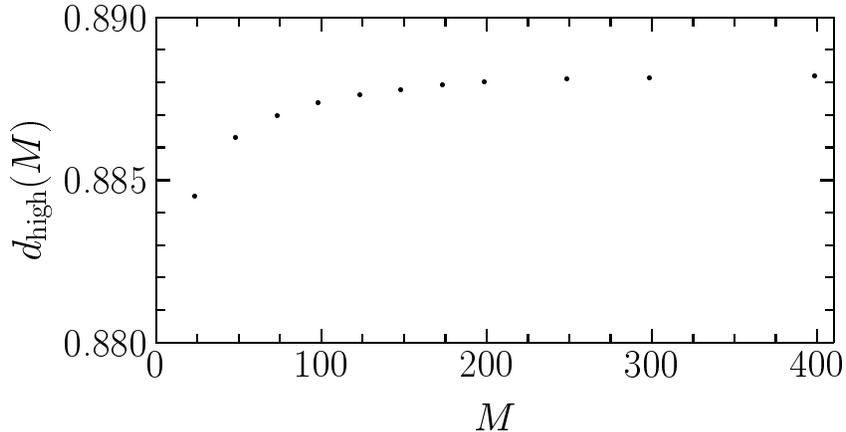,%
       bbllx=90pt,bblly=496pt,bburx=446pt,bbury=684pt,%
       width=11.0cm,silent=}
\]
\caption[fig:dhigh400]
        {\tenrm Determination of the exponent $d$ for the
        upper branch of the anomalous-dimensions spectrum.
\label{fig:dhigh}}
\end{figure}

The anomalous-dimensions spectrum can be reproduced by the logarithmic
fit
\begin{equation}
  \gamma _{n}(M)
=
  c + d \ln (M + b) \; .
\label{eq:gammalog}
\end{equation}
Up to order 150 both sectors of eigenvalues are included, i.e.,
$S_{n}=\pm 1$.
Beyond that order, for reasons of technical convenience, only the
antisymmetric ones have been taken into account.
However, since the spacings become very dense asymptotically, this poses
no restrictions on the validity of the calculation.
The upper envelope of the spectrum is best described by the following
values of the parameters with their errors:
$b=1.90989 \pm 0.00676$,
$c=-0.637947 \pm 0.000634$, and
$d = 0.88822 \pm 0.000119$.
For the lower envelope, the corresponding values are
$b=3.006 \pm 0.483$,
$c=-0.3954 \pm 0.0290$, and
$d = 0.59691 \pm 0.00545$.
The spacing of eigenvalues at very large order is reproduced by the
values $b=-0.027 \pm 0.728$,
$c=-0.2460 \pm 0.0248$, and
$d = 0.291883 \pm 0.00475$.
It is worth remarking that the previous logarithmic fit
$
 \gamma _{n}(M)
=
 \left[
       {\rm log}_{10} (2.13M + 1.4)
\right]^{1.48}
$,
given in \cite{THESIS97}, which takes into account all orders up to 150,
deviates from that in Eq. (\ref{eq:gammalog}) only by an amount less
than $5\%$ at order 400.
Hence, we conclude that the calculated spectrum shows already asymptotic
behavior, rendering the inclusion of still higher orders superfluous.
In particular, as one observes from fig. \ref{fig:dhigh}, the exponent
$d$ for the upper envelope shows scaling behavior for $M\geq
200$, approaching fast the value $0.8882$.
Physically, the logarithmic rise of the spectrum is due to the enhanced
emission of soft gluons, reflecting the fact that the probability for
finding bare quarks decreases \cite{DDT80}.

%

\begin{figure}
\tighten
\[
\psfig{figure=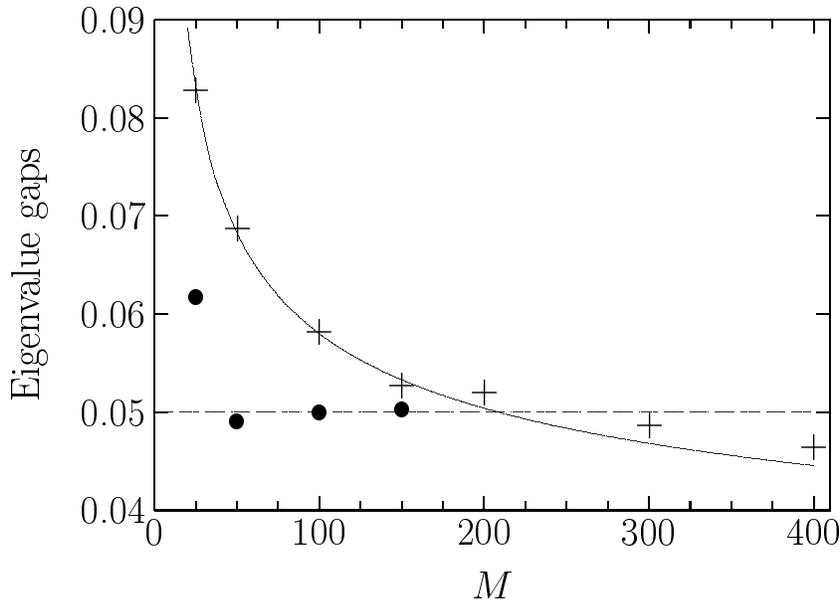,%
       bbllx=90pt,bblly=426pt,bburx=447pt,bbury=685pt,%
       width=11.0cm,silent=}
\]
\caption[fig:decoupling]
        {\tenrm Gaps of eigenvalues in the spectrum of anomalous
        dimensions.
\label{fig:gapgraph}}
\end{figure}

Let us now turn to the recent work by Braun et al. \cite{BDKM99}.
These authors claim that the lowest two eigenvalues of the spectrum
decouple, being separated from the others by a gap, which they interpret
as resulting from the formation of a diquark state.
Let us see whether our {\rm exact} large-order calculation confirms
their finding.
Fig. \ref{fig:gapgraph} shows the gap at each particular order between
the lowest two anomalous dimensions and the 3rd one as a function of the
order $M$ (marked by black dots) in comparison with the gap between the
7th and 8th anomalous dimensions (denoted by crosses), which we found to
be the largest gap after the 3rd one.
One observes that indeed beyond order 50, the gap of the two lowest
anomalous dimensions remains constant (up to the maximum considered
order 150), whereas the gap of the higher ones decreases logarithmically
like $0.267133/\ln M$ (solid line), where the first four shown values
were fitted.
Note that beyond order 200 only the antisymmetric values were used,
so that the calculated gaps in this sector lie slightly higher than
the gaps between the combined set of values of anomalous dimensions.
This behavior can indeed be interpreted as supporting the results of
\cite{BDKM99}, though our exact calculation gives for that gap a
somewhat higher value (cf. fig. \ref{fig:gapgraph}) than their
approximate calculation.

%

In summary, we have provided convincing evidence that the large-order
behavior of the anomalous-dimensions spectrum of trilinear twist-3 quark
operators with total derivatives tends to grow logarithmically with
order.
The exponent for the upper envelope of the spectrum was determined
with high precision and found to have the asymptotic value 0.8882.
The lowest two levels of the spectrum seem to be separated from higher
ones by a gap -- as found by Braun et al. \cite{BDKM99} -- possibly
indicating the formation of a binary quark system (diquark cluster)
inside the nucleon.
Yet, scepticism remains whether this finding will survive the inclusion
to the kernel of the nucleon evolution equation of higher-order
contributions which comprise gluon self-interactions.
We have presented explicit results for a set of 15 orthonormalized
eigenfunctions of the nucleon evolution equation, employing a
symmetrized basis of Appell polynomials, whereas still higher states can
be readily constructed by the means developed and used in this work.
These results suggest that the objections raised in \cite{BDKM99}
against the use of Appell polynomials in solving the nucleon evolution
equation are not really justified.

\bigskip
\ac
We would like to thank Dieter M\"uller, Pavel Pobylitsa, Maxim
Polyakov, and Prof. Grigoris Tsagas for useful discussions.
\newpage       

\newpage       
\end{document}